\documentstyle[times,epsfig]{mn}
\newbox\grsign \setbox\grsign=\hbox{$>$} \newdimen\grdimen \grdimen=\ht\grsign
\newbox\simlessbox \newbox\simgreatbox \newbox\simpropbox
\setbox\simgreatbox=\hbox{\raise.5ex\hbox{$>$}\llap
     {\lower.5ex\hbox{$\sim$}}}\ht1=\grdimen\dp1=0pt
\setbox\simlessbox=\hbox{\raise.5ex\hbox{$<$}\llap
     {\lower.5ex\hbox{$\sim$}}}\ht2=\grdimen\dp2=0pt
\setbox\simpropbox=\hbox{\raise.5ex\hbox{$\propto$}\llap
     {\lower.5ex\hbox{$\sim$}}}\ht2=\grdimen\dp2=0pt

\def\be{\begin{equation}}
\def\ee{\end{equation}}
\def\bc{\begin{center}}
\def\ec{\end{center}}
\def\beq{\begin{eqnarray}}
\def\eeq{\end{eqnarray}}

{

\begin{document}

\title{Electromagnetic Catastrophe in Ultrarelativistic Shocks and 
the Prompt Emission of Gamma-Ray Bursts}

\author[Boris E. Stern] 
{\parbox[]{6.8in}{Boris E. Stern$^{1,2}$\thanks{E-mail:
stern@lukash.asc.rssi.ru}
}\\
$^{1}$Institute for Nuclear Research, Russian Academy of Sciences,
Moscow 117312, Russia\\
$^{2}$Astro Space Center of Lebedev Physical Institute,
Moscow, Profsoyuznaya 84/32, 117997, Russia}

\date{Accepted, Received}

\maketitle

\begin{abstract}
 
 It is shown that an ultrarelativistic shock with the Lorentz factor
of order of tens or higher  propagating in a moderately 
dense interstellar medium (density above $\sim 1000 $\ cm$^{-3}$) 
undergoes a fast dramatic transformation into a highly radiative state.  
The process leading to this phenomenon resembles the first order Fermi acceleration with the difference
that the energy is transported across the shock front by photons  
rather than protons. The reflection of the energy flux crossing the shock front in both directions 
is due to photon-photon pair production and Compton scattering.    
Such mechanism initiates a runaway nonlinear pair cascade fed directly by the kinetic energy
of the shock. Eventually the cascade feeds back 
the fluid dynamics, converting the sharp shock front into a smooth velocity gradient 
and the runaway evolution changes to a quasi-steady state regime. 
This effect has been studied numerically using the nonlinear Large Particle Monte-Carlo 
code for the electromagnetic component and a simplified hydrodynamic description of 
the fluid. 
 The most interesting application of the effect is 
the phenomenon of gamma-ray bursts where it explains a high radiative efficiency
and gives a perspective to explain spectra of GRBs and their time variability.
The results predict a phenomenon of ``GeV bursts'' which arise if the density 
of the external medium is not sufficiently high to provide a large compactness.

\end{abstract}



\section{Introduction}

 Astrophysical objects where we observe (active galactic nuclei) or infer 
(gamma-ray bursts, soft gamma-repeaters) 
a relativistic bulk motion demonstrate surprisingly high gamma-ray 
luminosity. In some cases the energy release in gamma-rays 
can be comparable to the total kinetic energy of the bulk motion in the source.
This implies a very efficient mechanism of the energy conversion 
of the bulk motion into the hard radiation. It is clear that in the case of 
explosive phenomena like GRBs the radiation originates from ultrarelativistic 
shocks (Cavallo \& Rees 1978; Rees \& M\'esz\'aros 1992). This is less certain 
in the case of AGN jets, however there exist 
a large number of works considering shocks as the main source of AGN jet 
radiation (for a review see Kirk, 1997).    

One can describe the energy stored in the shock in two ways: as a total 
kinetic energy of ejecta and shocked matter or as their internal energy.  
There are known two different ways of shock energy release:
the first order Fermi acceleration and different mechanisms of 
internal energy dissipation. Fermi acceleration does not require 
the dissipation as it is fed directly by the kinetic energy of the bulk 
motion.
On the other hand, Fermi acceleration has a number of limiting factors 
in available power and radiative efficiency (see, e.g., Bednarz \& Ostrovsky 1999) .
With this reason Fermi acceleration is usually considered as a source of 
high energy cosmic rays while the intensive radiation is traditionally 
explained in terms of the internal energy dissipation. 

 This work suggests a highly radiative analog of Fermi acceleration
where the main role is played by photons rather than by 
charged particles. A similar idea about important role of neutral
particles in shocks has been suggested independently in a very recent paper 
by Derishev et al. (2003). 
For a simple illustration the shock and the external 
medium can be represented
as two mirrors moving with an ultrarelativistic velocity  respectively to each
other. Let $\xi_s$ and $\xi_e$ be their reflection coefficients (depending on 
the kind and the energy of a particle) and $\Gamma$ -- the Lorentz factor of the 
shock. A photon being elastically reflected head-on from a moving mirror 
gains a factor $ \sim \Gamma^2$ in its energy. 
Thus the total energy of photons 
participating in back and forth motion between the mirrors can be roughly
described by the equation 
\begin{equation}
dE/dt = (\xi_s \xi_e \Gamma ^2 - S)E/t_c
\end{equation}
where $t_c$ is a characteristic time of reflection cycle  and $S$
is the probability of the 
particle escape. Actually equation (1) represents not more than an 
instructive toy model since a real situation  
is essentially non-local ($t_c$ is uncertain) and require an integration through particle 
spectra. If $\xi_s \xi_e \Gamma^2 > S$, 
the total energy of particles will grow exponentially until mirrors 
decelerate.

 In the case of the first order Fermi acceleration the reflection 
results from the diffusion of supra-thermal charged particles 
in magnetic field. 
 How this two-mirror approach will look when applied to photons? 
 If a soft photon (e.g. of a synchrotron nature) interacts only with 
electrons, the reflection coefficients 
$\xi_s$ and $\xi_e$ are of order of the Thomson optical depth of each medium
($\xi_s \sim \xi_e \sim \tau_T$, see section 2). 
 This case does not differ essentially from the well known thermal
Comptonization in relativistic regime at electron temperature $\Gamma-1$. 
 The photon boosting terminates 
as soon as the photon gains a sufficient energy to interact in Klein-Nishina 
regime when reflection coefficients decrease. The Compton boosting is
probably not very important by itself, however it could provide photons
above pair production cutoff and give rise to a new much more powerful 
effect. 

 The importance of pair production in GRB shocks and AGN jets has been 
emphasized in many works, e.g., by Thompson (1997), Ramires-Ruis et al. (2000)
and M\'esz\'aros, Ramires-Ruis \& Rees (2001).
Thompson (1997) and Ramires-Ruis et al. (2000) stated that the pair production can enhance 
the radiation 
efficiency of the shock providing new supra-thermal particles as seeds for 
Fermi acceleration. In this work we show that the pair production plays
a crucial role initiating a runaway energy release.

 A necessary condition to trigger this process is the existence of 
a seed radiation field in the vicinity of the shock. This radiation should be 
dense in terms of the photon number: $n_p  R \sigma_T \gg 1$ where $R$ is a
characteristic scale, both $n_p$ and $R$ correspond to the comoving frame of 
the shock where photons are approximately isotropic. Such photons can appear 
e.g. as the result of an internal energy dissipation and undergo the Compton boosting.

Now let a photon of energy 
$\varepsilon \gg 1$ (hereafter $\varepsilon$ is the photon energy in electron mass units) 
moves upstream across the shock front. Then, interacting with a soft photon, it produces 
an e$^+$e$^-$ pair in unshocked matter ($\xi_e \sim 1$ in this case).
The pair is bound to the rest frame of the external matter if the e$^+$ and e$^-$ 
Larmor radii do not exceed the scale characterizing the problem. Particles 
are advected downstream across the shock front, losing a fraction of their energy
 to Comptonization. Some of downstream Comptonized photons can
produce new pairs behind the shock front which in turn upscatter soft photons upstream.
So we deal with multiple electromagnetic interactions and the whole process can be 
characterized as an electromagnetic cascade fed by kinetic energy of two converging 
media or {\it cascade boosting}. 
 The energy involved in the cascade can grow exponentially 
and eventually the process can reach a nonlinear stage when the 
boosting rate increases.

 All above conclusions have been made in assumption that the shock front separating 
unshocked and shocked media is very sharp and can be characterized as a viscous jump.
It is known from studies of nonrelativistic shocks that the viscous jump can disappear
transforming into a smother transition due to radiation pressure ahead the shock.
Thompson \& Madau (2000) have studied preacceleration of external matter for the 
case of a GRB external shock and demonstrated that the pair production (pair loading)
is extremely important in this respect and feeds back the radiation efficiency.
Beloborodov (2002) has considered a more extreme case when the radiation of GRB ejecta 
sweeps up the external medium far ahead the shock. In that case the energy  
should be supplied from internal shocks. Our case is closer to that of Thompson \&
Madau (2000). The preacceleration of external medium transforms the viscous jump  
into a smoother velocity gradient.  This transformation removes the radiative
divergence: now an upstream photon most probably interacts in preaccelerated 
medium and the energy gain in the interaction becomes much less than $\Gamma^2$. 
We should expect 
the formation of some steady state regime where the fluid velocity pattern is 
adjusted in 
a way to satisfy the energy conservation law. The luminosity at this stage 
should be close to its ultimate value $L \sim \Gamma^2 dm/dt$ where $m$ is the mass of 
the external matter swept up by the shock.

 Summarizing, we can expect the following evolution: 
(i) a seed radiation due to dissipation of the internal energy; 
(ii) Compton boosting and the cascade boosting of the
electromagnetic component in linear exponential regime;
(iii) a sharper nonlinear evolution when the boosted high energy component contributes 
to the soft target photon field;
(vi) transformation  of the shock front and setting up a highly luminous quasi-steady 
state regime. The entire event is so dramatic that it deserves the name 
``electromagnetic catastrophe''.     

 This qualitative illustration is confirmed below at a quantitative level.
The main tool in this study is the large particle nonlinear Monte-Carlo (LPMC) code
described in Stern et al. (1995).

 While the mechanism can work in different astrophysical phenomena
this study is focused on its application to GRBs. Correspondingly,
the explored region of parameters represents the prevailing
view on GRBs (large Lorentz factors, moderate densities).

 In Section 2 we describe the general formulation of the problem, 
the accepted assumptions and the range of parameters. 
Section 3 briefly describes the approach to the numerical simulations of the effect.
Section 4 presents the results of simulation runs  
demonstrating the  evolution of the system from a low luminosity 
initial state through the catastrophe to a highly radiative post-catastrophe regime. 
Section 5 extends the range of the phenomenon to lower densities and outlines the 
threshold where the catastrophe is possible at certain assumptions.
Section 6 describes a phenomenon of GeV bursts which should arise
when the compactness is insufficient to provide optically thick pair loading.
Possible associations with existing data are discussed. In section 7 we 
discuss various GRB scenario associated 
with electromagnetic catastrophe and the issue of the GRBs time 
variability.

\section{Formulation of the problem}

\subsection{The simplified description of the shock}

 For convenience we consider the shock in the comoving system which is preferable
for its higher symmetry: the 
density contrast is moderate (factor 4 in an idealized case), upstream and 
downstream energy fluxes can be comparable. More explicitly, we bound 
our reference
system to the contact discontinuity, its Lorentz factor hereafter is $\Gamma$.
The shock front in this frame moves upstream with semirelativistic velocity 
$\beta_s$ (hereafter, $\beta$ is the velocity in units of the velocity of 
light). We assume that the shocked external matter between the shock front and 
the contact discontinuity is of a constant density and in rest relatively to 
the contact discontinuity. Actually this region can be subject to a 
semirelativistic bulk motion (perhaps with a turbulent component) 
depending on the distribution of external matter and shock dynamics. 
Neglecting such motion we remove an additional source of the free energy 
and since we are interested in an explosive energy release this is a 
conservative assumption. 

 We do not know and do not consider the state of 
ejecta behind the contact discontinuity, prescribing to it a role of a 
heavy piston, carrying the main fraction of the kinetic energy. It is assumed that ejecta 
does not emit or reflect 
radiation. Actually ejecta can contribute the seed radiation and the 
downstream-upstream reflectivity.   

 We assume a spherical symmetry at least within 
a cone of opening angle $1/\Gamma$ respectively to the line of sight. 
Due to relativistic effects we almost do not observe the emission 
of the 
shock beyond this cone. Events in the shock separated by angle more than $1/\Gamma$ are 
not causally connected since the acceleration stage. 
Most probably the ejecta is strongly beamed within a solid angle $\Omega \ll 1$, 
we exclude $\Omega$ from the consideration using the isotropic equivalents 
for global energy parameters: the kinetic energy of ejecta, $E$, and the 
luminosity, $L$. The actual energetics is then described by $ \Omega E$
and $ \Omega L$. 

 The spherical shock viewed from the comoving system is parabolically curved
and the matter moves outward from the line of sight, relativistically at 
the angular displacement $\sim 1/\Gamma$ from the line of sight. For 
simplicity we adopt one-dimensional slab geometry with no
bulk motion in the axial direction. 
This is not a conservative simplification  
because in this way we violate kinematic and causality constraints on interactions of photons
emitted from distant regions of the shock. 
In order to avoid this problem, the consideration was restricted to a 
comparatively small time 
(less than a half deceleration time) when the above constraints are still not important.

 The unshocked interstellar matter (ISM) in the comoving system is a 
cold fluid with bulk Lorentz factor $\Gamma$. The ordered particle motion  
dissipates into chaotic motion with the same average thermal Lorentz factor 
$\Gamma$ when the fluid crosses the shock front. The energy density 
contrast across the front and the front velocity in the 
comoving system, $\beta_s$, result from the energy conservation:
\begin{equation}
 U_s \beta_s = U_{\rm o} (\beta + \beta_s),
\end{equation}
where  $U_s$ is the energy density of the shocked matter,$\beta$ is the 
external fluid velocity, $U_{\rm o} =
\Gamma^2 \rho m_p c^2$, $\rho$ is the rest frame ISM number density, and
from the pressure -- momentum balance:
\begin{equation}
 P_s = {1 \over  3} U_s = U_{\rm o} \beta (\beta + \beta_s).
\end{equation}
 In the ultrarelativistic limit ($\beta = 1$) equations yield $\beta_s = 1/3$
$U_s = 4 U_{\rm o}$. For a more detailed description of a relativistic 
shock see Blandford \& McKee (1976). 

 The thickness of the shock (between the shock front and the contact discontinuity)
at the constant external density and spherical geometry is $\Delta X = {1 \over 12} R/\Gamma$
in the comoving system.

\subsection{Main parameters and dimensionless units}
 
 External magnetic field $H_e$ is different for cases of the stellar wind dominated and the ISM dominated environments.
In the first case, the field is predominantly radial and processes leading to the catastrophe 
are inhibited ($\xi_e$ is very low) unless the field is affected by radiation front ahead the shock.
Most probably it must be affected by the two stream instability generating a chaotic component.
Moreover, a spherical asymmetry of the environment and ejecta together with the charge asymmetry  
(Compton scattering on electrons) can induce a large scale transversal field
similar to that induced in an atmospheric nuclear explosion. 
 
In the ISM dominated case, the field originally should have a large scale geometry and a 
substantial transversal component.
In this work only the ISM case is studied quantitatively, note that the interaction of the shock
with a slow shell of matter ejected by a GRB progenitor essentially does not 
differs from the ISM case. 
ISM magnetic field depends on the external density $\rho$ and the reasonable assumption is this case is $H_e \sim 10^{-5}
\sqrt \rho$ G, where $\rho$ is in cm$^{-3}$ units. Hereafter we treat $H_e$ as the transversal component.

 The lower limit for the magnetic field in the shock $H$ follows from Lorentz transformation 
and the compression by factor 4: $H > H_g = 4 H_e \Gamma$. However, this field can be much stronger being 
generated at the shock. At this step we can rely on numerical study of the two stream instability by 
Medvedev \& Loeb (1999). According to their simulations the ensured value of energy density of 
generated field, $U_B$, is $\sim 0.01 - 0.1$ of electron energy density in the shock, $U_e$,
and, tentatively, the field can be amplified up to
$U_B \sim 0.01 - 0.1$ of proton energy density. 

The parameters of the problem are: initial Lorentz factor $\Gamma$, total 
kinetic energy $E$, 
external number density $\rho$, characteristic rest frame size $R$, 
seed luminosity of soft 
photons $L_s$, their spectrum, values of magnetic field in both regions, 
$H$ (shock) and $H_e$.
As a reasonable hypothesis for the seed spectrum we accept a power law 
$dN/d\varepsilon \propto \varepsilon^\alpha$, in the range $\varepsilon_1 < \varepsilon <\varepsilon_2$
(see \S 2.3).

 In the case of GRBs the isotropic kinetic energy of ejecta can vary within two orders of 
magnitude or more.
Maximal observed hard X-ray -- gamma-ray GRB energy release is $\sim 5 \cdot 10^{54}$erg.
We accept $E = 10^{54}$ erg as a baseline. 
In this work we study the case of a constant external density which is a reasonable 
starting point. The scale of the problem can be 
characterized by the deceleration radius (Rees \& M\'esz\'aros 1992):
\begin{equation}\label{eq:dr}
R_d = \left[E/(m_p c^2 {4 \over 3} \pi \rho \Gamma^2)\right]^{1/3} = 0.24 \cdot 10^{16} 
E_{54}^{1/3} \rho_{6}^{-1/3} \Gamma_{2}^{-2/3} {\rm cm}
\end{equation}

 In the spherical geometry, the scale changes as  the shock 
propagates, therefore we adopt $R = 0.5 R_d$ which approximately 
describes the shock propagation between $0.5 R_d$ and $R_d$.
Using the rest frame scale $R$ we can define dimensionless distance, $x$,
and time, $t$, 
to work within the comoving system: $x= X/(R/\Gamma) $, $t= T/(R/\Gamma c)$,
where $X$ and $T$ are comoving distance and time.
Note that the allowed working range is $x < 1$ and $t < 1$. 
The time in observer frame is $T_{\rm obs} = t R /(2 \Gamma^2 c)$.

It is convenient to describe the energy budget in terms of dimensionless 
compactness. The latter defines the importance of pair production and,
more generally, the level of nonlinearity of the system. The simplest way
to introduce this parameter is via the energy column density through 
the system normalized to the electron mass and multiplied by Thomson cross 
section: 
\begin{equation}
\omega = {U \over m_e c^2} \sigma_T {R \over \Gamma},
\end{equation}
 where $U$ is an energy density in the comoving system.
Particularly, the magnetic compactness is:
\begin{equation}
 \omega_B = {H^2 \over 8\pi m_e c^2} \sigma_T {R \over \Gamma}.  
\end{equation}
 This is the compactness describing the energy content of the system. Traditionally the compactness 
is expressed through the luminosity of the system $L$: 
\begin{equation}
\ell = {L \sigma_T \over R m_e c^3}.
\end{equation}

The rest (observer) frame compactness for the ultrarelativistic shock is 
meaningless since the radiation is 
strongly collimated. The parameter has the stated above physical meaning only in the shock comoving 
system where particles have a wide angular distribution. Assuming the comoving size of the 
system $R/\Gamma$ (with the transversal size of the same order, see \S 2.1) and substituting $L$ by the 
luminosity flux: $L \sim F  (R/\Gamma)^2$,
with the accuracy of the order of $\pi$ one gets:
\begin{equation}
\ell = {F \over m_e c^3} \sigma_T {R \over \Gamma}.
\end{equation}
The total energy budget in the comoving frame
constitutes the kinetic energy flux of the fluid crossing the shock front. 
The corresponding dimensionless expression,
kinetic compactness, is:
\begin{equation}
\ell_{\rm o} = \rho \Gamma^2 m_p c^3 {\sigma_T \over m_e c^3} {R \over \Gamma} = \tau_T {m_p \over m_e} \Gamma.
\end{equation}
 Hereafter we denote energy compactness as $\omega$ and the luminosity
(power) compactness as $\ell$, and $\ell = d\omega /dt$.  
 
\subsection{Seed radiation}

 The seed radiation in the shock can be contributed by (i) the synchrotron -- self-Compton    
radiation of shocked electrons which have the thermal Lorentz factor 
$\Gamma$; (ii) other channels of internal energy dissipation (e.g. dissipation 
of magnetic field, plasma waves, turbulence); 
(iii) external (partially side-scattered) photons from 
the fireball and a progenitor star; (iv) high energy 
seed photons associated, e.g, with Fermi acceleration and photo-meson production.  

 The most confident source is (i) unless synchrotron radiation of shocked 
electrons is strongly self-absorbed. Its fraction in the total energy budget 
is limited by the factor $m_e/m_p$.
 At some parameters, this seed radiation is sufficient to
start the evolution leading to the catastrophe (see Section 3). In such case we deal with the most 
conservative ``minimal hypothesis''. At other parameters the minimal hypothesis is insufficient for the catastrophic evolution.
Then we have have to consider additional channels of internal energy 
dissipation (ii)
or external photons (iii) and (iv).
 We would like to avoid a serious consideration of corresponding mechanisms
in this work and use a very moderate {\it ad hoc} hypothesis of the seed soft
photon field.  

As a baseline, we accept 
the compactness of such emission 
$\ell_s \sim \tau_T \Gamma$ which is only $m_e/m_p$ fraction of the total 
energy budget.         
The model spectrum of this seed radiation has been taken as a power law fragment 
with $-2.5 < \alpha < -1.5$ (a free parameter) in the range between $\varepsilon_1$ 
(typically $\varepsilon_1 \sim 10^{-8}$ 
which is close to the self-absorption cutoff at $H \sim 100$ G) and $\varepsilon_2$ (a free parameter). Such weak seed radiation is sufficient to give rise 
to the catastrophe in the most of trials described in this work. The situations
when the effect implies a stronger seed radiation are discussed in section 7.

\subsection{Reflection of the energy flux}

 Reflection of a particle flux from the external environment and from the shock 
is of a primary importance for the cascade boosting (see Eq. 1). Here we consider 
the reflection of a high energy component of the electromagnetic cascade 
where the direct Compton 
reflection of photons is inefficient. In this case the reflectivity of the medium is  
due to photon-photon generated pairs gyrating in the magnetic field.
The requirement that an electron can change its direction to the opposite 
without a substantial energy loss is a serious limiting factor for the boosting 
rate.

 In the case of the first order Fermi acceleration the main process
responsible for the particle scattering upstream across the shock is 
the diffusion in a chaotic magnetic field (for a recent review 
see Gollant 2002). In our case the charged particle does not need to cross
the shock, it is sufficient if it just changes its direction upstream. Then
photons Comptonized by the particle moving upstream can cross the shock. 
The deflection of a particle from downstream to upstream hemispheres is 
a much more probable event than its diffusion upstream the shock, required 
for Fermi acceleration. Indeed, in the comoving frame of the shocked matter 
the shock front moves ahead semirelativistically, with $\beta_s= 1/3$, and a 
charge particle has to diffuse upstream faster.

 The downstream -- upstream reflection of a particle can proceed in the 
following ways:

(i). If the Larmor radius is less than the correlation length of the magnetic field,
the particle turns at a half Larmor orbit. This can be the case for particles of a 
moderate energy. 

(ii). If the Larmor radius exceeds the correlation length, 
the particle diffuses in the chaotic field. 

(iii). Except the chaotic field $H$ there could 
exist a large scale component, resulting from a large scale external magnetic field due to 
the flux conservation: $H_g \sim 4 H_e \Gamma$. Then a high energy particle can turn 
around at a half Larmor orbit in the field $H_g$.
 Processes (ii) and (iii) are competing and the leading one depends on
concrete parameters. In this work only the effect of the
global component has been taken into account since the latter should exist in 
the adopted model of the external environment. The effect of the diffusion 
is the matter for future studies.

 The reflection efficiency depends on the fraction of energy which the particle 
loses before it turns around upstream. The energy loss rate is 
described by
\begin{equation}
-d\gamma /dX = \left({H²\over 8\pi}+U_{\rm ph} \right) {\sigma_T \over m_e c^2} \gamma^2 = C \gamma^2, 
\end{equation}
where $X$ is the comoving distance.
 The equation yields 
\begin{equation}\label{eq:loss}
\gamma(X) = \gamma_{\rm o} /(1+ \gamma_{\rm o} X C),
\end{equation}
 where 
\begin{equation}
C = H^2 (1+U_{\rm ph}/U_B) 3.3 \cdot 10^{-20} {\rm cm}^{-1}.
\end{equation}
where $U_B$ and $U_{\rm ph}$ are energy densities of magnetic field and soft radiation 
field respectively. 
The term $U_{\rm ph}/U_B$ describes the relative contribution of Compton losses.
From equation (\ref{eq:loss}) one estimates the maximum energy of downstream
-- upstream reflected 
particle  $\gamma_{\rm max} \sim 1/XC$. Substituting X by the half Larmor orbit 
\begin{equation}\label{eq:lr}
\pi R_L =  5.3 \cdot 10^3  {\gamma_{\rm max} \over H_g} cm,  
\end{equation}
one obtains the final expression for the maximal 
energy of a reflected particle:

\begin{equation}\label{eq:gmax}
 \gamma_{\rm max} = 0.75 \cdot 10^8 {H_g^{1/2} \over H} (1+U_{\rm ph}/U_B)^{-1/2}. 
\end{equation}

 The downstream -- upstream reflection is inefficient if $U_{\rm ph}/U_B \ll 1$.
Then the main energy of an electron is emitted as a relatively soft 
synchrotron radiation rather than a hard Comptonized photons.
The limiting energy of synchrotron photons emitted by an electron 
after half Larmor orbit is $\varepsilon \sim 200$ at $H_g \sim H$.
This energy is still sufficient for pair production, however 
the cascade boosting in this case will be substantially slower and the
compactness threshold for the electromagnetic catastrophe will be higher.
In this work we study the case of low $H$ and weak dissipation.
If the generated field is much stronger but its dissipation into 
electromagnetic component is fast, then we have $U_{\rm ph}/U_B \sim 1$ again.
The boosting is inhibited if the field is strong and its  
dissipation is slow ($U_{\rm ph} \ll U_B$) .    

 In the case of the external environment (upstream-downstream reflection) the  
electron synchrotron losses  
are small. Indeed, in the rest frame equation (\ref{eq:gmax}) with 
$H_g = H = 10^{-5}\sqrt \rho$ yields
\begin{equation}
\gamma_{r,{\rm max}} = 1.3 \cdot 10^{10} \rho_6^{-1/2} (1+U_{\rm ph}/U_B)^{-1/2} 
\end{equation}
 where the subscript $r$ refers to the rest frame
and $\gamma_{\rm max} \sim \Gamma \gamma_{r,{\rm max}}$. 
The term $U_{\rm ph}/U_B$ is large in the external environment, (i.e. 
Compton losses are much larger). However, Comptonized 
photons have predominantly downstream direction and contribute 
to the cascade boosting.

The main limiting factor in upstream -- downstream reflection is
the requirement that the Larmor radius should be smaller than the  
characteristic size of the problem. Actually, the upstream particle can be 
advected back across the shock front when it deflects bu just the angle 
$1/\Gamma$. In this case it has a much less energy gain than $\Gamma^2$.
Both effects were accounted for in the numerical simulation. 
 From the limit on the gyroradius $R_L < R$ equations (\ref{eq:dr}) and 
(\ref{eq:lr}) yield: 
\begin{equation}\label{eq:gmaxe}
\gamma_{r,{\rm max}} < 1.4 \cdot 10^{12} H_e  E_{54}^{1/3} \rho_{6}^{-1/3} \Gamma_2^{-2/3}.
\end{equation}
 Using the hypothesis $H_e = 10^{-5} \sqrt{\rho}$ and transforming to the comoving frame one gets:
\begin{equation}
\gamma_{\rm max} < 1.4 \cdot 10^{11} E_{54}^{1/3} \rho_{6}^{1/6} \Gamma_2^{1/3}.
\end{equation}

 Note that the limits (\ref{eq:gmax}) and (\ref{eq:gmaxe}) have an opposite dependence on 
the magnetic field. A faster boosting takes place in the case of a stronger external
field and a weaker field in the shock. The above estimates are given for the orientation
while the numerical simulation reproduces directly the particle trajectory 
and the energy loss along it. 
The quantitative effect of these parameters on the 
boosting rate and examples of upstream and downstream particle spectra are demonstrated in  Section 5.

\section{Monte-Carlo simulations}
 
 The entire problem outlined above is too difficult for an analytic treatment.
In this work it is studied numerically using a Large Particle Monte-Carlo code 
(LPMC) developed by Stern (1985) and Stern et al. (1995). The code is
essentially nonlinear: the simulated particles constitute
at the same time a target medium for other particles. Large particle (LP) method 
in this application means that each simulated particle represents $w$ real 
particles (for GRBs $w > 10^{50}$). The most reasonable weighting scheme
except some special cases is to set $w \varepsilon = const$, where $\varepsilon$ 
is the energy of the particle and the constant can vary since the
total energy involved in the simulation can change. The number of LPs 
was $2^{17} = 131072$.

The version of LPMC
used here treats Compton scattering, synchrotron radiation,
photon-photon pair production, and pair annihilation. Synchrotron self-absorption
was neglected as it consumes too much computing power and is not very important
in this application. All these processes are 
reproduced without any simplifications at the microphysics level. 
On the other hand, a number of serious macroscopic simplifications
has been done.

 Since the number of large particles which is limited by the computing power 
available 
at the moment did not allow full three-dimensional simulation, the problem 
was reduced to one dimension, $x$, along the shock propagation. 
While locations and momenta of LPs were three dimensional, the target density 
was averaged over slab layers and the fluid representation was one-dimensional.

The trajectories of electron and positrons in the magnetic field were 
simulated directly assuming transversal geometry of the field $H_g$ 
in the shock and $H_e$ in the external matter. Thus the problem of the 
reflection qualitatively discussed in \S 2.4 was implemented in the
numerical simulation. 
Protons were assumed to be cold in the fluid frame unless they have crossed the 
dissipative shock front. 

 Hydrodynamic part of the problem becomes very difficult since the
energy and momentum transferred by LPs to the fluid fluctuate.
Attempts to simulate the fluid with internal pressure have led to rising 
instabilities
at semirelativistic velocities. Therefore a dust approximation (where the 
internal fluid pressure is neglected) has
been adopted and the fluid velocity behind the shock front was artificially fixed
to zero. 
  
 The dust approximation works satisfactory until the fluid has a moderate 
temperature and energy density in its comoving frame (i.e. when it is 
ultrarelativistic in the shock comoving frame). This circumstance was used by 
Beloborodov (2002) and Thompson \& Madau (2000) who also described the 
preacceleration of the external medium by the radiation front 
in terms of dust approximation. The approximation does not work at all 
in the shock where the pressure at a relativistic temperature is
comparable to the radiation pressure. The adopted hydrodynamic anzats
allows to describe the effect in general and fails to account for possible 
interesting effects associated with semirelativistic motion behind the 
radiation front. The approach can be considered as a ``zero approximation''.

\section{Evolution of the system through the catastrophe}

 To illustrate qualitative arguments discussed in Introduction with 
numerical simulations, we present results of two runs with different 
parameters. First, for Lorentz factor $\Gamma = 100$ and the density of 
external environment $\rho = 3 \cdot 10^5$ cm$^{-3}$. Second, for $\Gamma=30$, 
$\rho = 10^6$cm$^{-3}$. Other parameters for these cases are $R = 1.9 \cdot 10^{15}$ cm,
$\ell_{\rm o} = 63$, $T_{\rm o} = 3.1$s and $R=2.8 \cdot 10^{15}$cm, 
$\ell_{\rm o} = 94$, $T_{\rm o} = 52$s.
 The first case is closer to the traditionally assumed
value of $\Gamma$ for GRBs, however it is more difficult 
for numerical treatment since statistical fluctuations associated with finite number of 
large particles being amplified by factor $\Gamma^2$. With this reason the first run 
tracks the evolution of the system only till the beginning of the 
post-catastrophe stage.  

 Initial state is empty of photons and consists only of fluid of cold protons
and electrons with bulk Lorentz factor $\Gamma$, representing external medium viewed from 
the comoving system. Seed photons appear as the (partially self-absorbed) 
synchrotron radiation of shocked electrons. 
In this case, the self-absorption is not very strong
and can be ignored.  The Comptonization of the synchrotron peak gives second 
and third peaks at higher energies (further Comptonization is inefficient 
because it proceeds in Klein-Nishina regime). These peaks are visible in Figure 1
at $t=0.3$. The Comptonized photons are produced both by shocked and external 
electrons.
The latter cause an anisotropy: the second peak is stronger in downstream 
photons
(Figure 1, lower panel), while the third peak is slightly more intensive 
upstream (Figure 1, upper panel).

\begin{figure}
\centerline{\epsfig{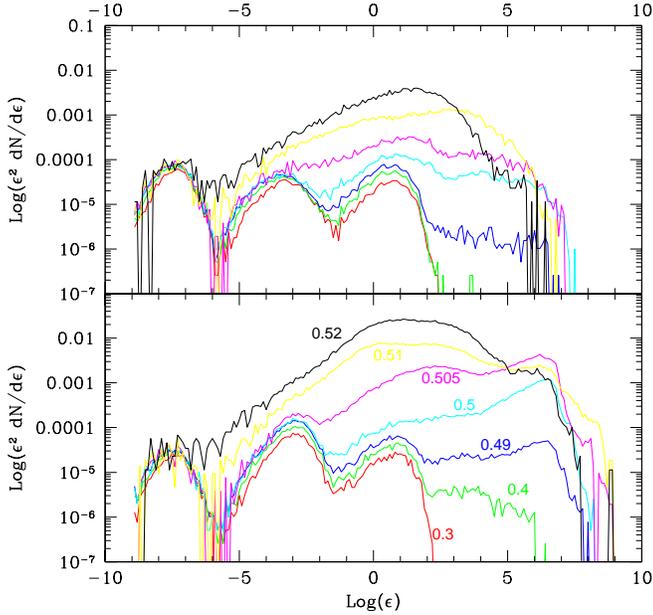}}
\caption{Instantaneous spectra of photons for $\Gamma = 100$, $\rho = 3 \ 10^5$cm$^{-3}$
during pre-catastrophe and catastrophe stages. Upper panel shows the 
upstream and lower -- the
downstream photon spectra.
Labels in the lower panel
indicate the dimensionless time from the start of the simulation. 
Curves in the upper panel correspond to the same time sequence.
Time unit corresponds  to 3.1 s in the observer frame.}
\end{figure}

At $t = 0.4$ a  high energy tail appears due to pair production by photons of the third peak
in the external medium (which provides comoving energy gain by factor $\Gamma^2$).
Then, around $t=0.5$, the evolution accelerates dramatically, the total energy of photons
rises by more than 3 orders of magnitude (the power supplied to photons jumps by factor
$\sim m_p/m_e$), and the spectrum transforms from harder to softer one. The evolution 
from $t=0.49$ to $t = 0.52$ will take about 0.1 second from the point of view of an 
external observer.
The simulation has been terminated at $t=0.56$ because numerical problems associated with
the high Lorentz factor become further serious. These problems are not 
principal and can be overcome just with the increase in the computing power.
The evolution of the system with time is shown in Figure 3.

 This example does not require any additional source of energy except the bulk kinetic 
energy of the fluid and any external photons. In this sense the case can be 
interpreted as a ``minimal model'', however one still has to assume a non-radial 
external magnetic field.

 The case of $\Gamma = 30$, $\rho = 10^6$cm$^{-3}$ does require additional assumptions.
First of all, the synchrotron radiation of electrons with $\varepsilon = 30$ is
deeply self-absorbed. Therefore an internal energy source accelerating a fraction of shocked 
electrons up to $\varepsilon \sim 200 - 300$ is required. The required power is, 
nevertheless,  much less than the total energy budget: $\ell_s \sim m_e/m_p \cdot \ell_{\rm o}$.
To avoid extra details we just ignored the self-absorption since the nature and energy 
of seed photons is not very important. Another problem is that the energy of 
the second
Comptonization peak is smaller than in the previous example and the catastrophe
develops too slowly. In section 5 it is shown that a much faster evolution can be induced 
by seed high energy photons which population undergoes an exponential breeding.
In this run a portion of upstream high energy photons with total 
energy compactness $\omega_s \sim 10^{-5}$ (which is $ \sim 10^{-7}$ of the
total energy budget) has been injected at $t = 0.25$.
Then the catastrophe has followed at $t \sim 0.3$.

\begin{figure}
\centerline{\epsfig{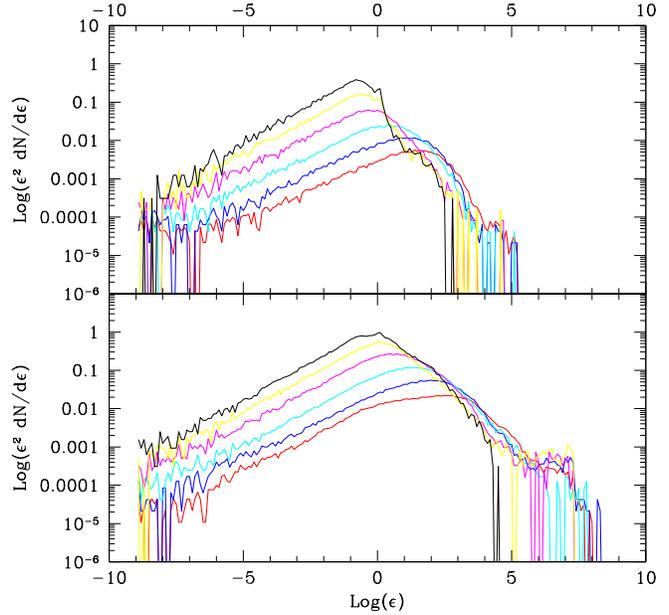}}
\caption{Instantaneous spectra of photons for $\Gamma = 30$, $\rho =  10^6$cm$^{-3}$
during catastrophe and post-catastrophe stages. Upper panel shows the upstream and lower shows the 
downstream photon spectra. 
The dimensionless time, from lower to upper curves, is: 0.285, 0.29, 0.31,
0.34, 0.4, 0.5, 0.67. Time unit corresponds 
to 52 s in observer frame.
}
\end{figure}

\begin{figure}
\centerline{\epsfig{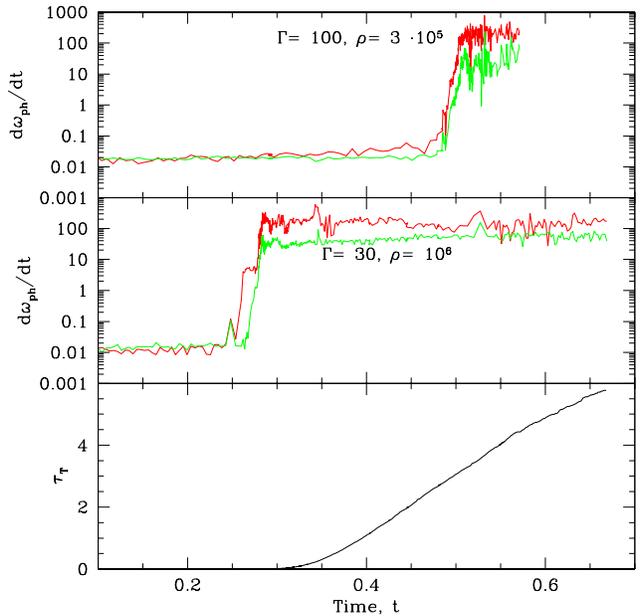}}
\caption{The evolution of the electromagnetic component with time.
Upper panel shows the time derivative of the total dimensionless energy 
$d\omega_{u}/dt$ of upstream (lower curve)
and downstream $d\omega_{d}/dt$ (upper curve) photons for $\Gamma = 100$, 
$\rho = 3 \ 10^5$cm$^{-3}$.
Middle panel: the same for $\Gamma = 30$, $\rho =  10^6$cm$^{-3}$. 
Lower Panel shows the 
evolution of pair Thomson optical depth for$\Gamma = 30$, $\rho =  10^6$cm$^{-3}$ . 
}
\end{figure}

 Figure 2 shows post-catastrophe photon spectra which can be treated as a continuation
of the spectra set presented in Figure 1 to later ages. Figure 4 shows the spatial 
distribution of main
quantities: the energy density of upstream photons, the pair number distribution and 
the bulk Lorentz factor. At the moment of the catastrophe the main fraction of photon
energy is tightly concentrated to the fluid velocity jump which has a step-like character
(see Figure 4, upper and middle panels, curve at $t=029$).
The reason is that the evolution at that moment is so fast, that abruptly 
released photons have no time to disperse wider. Then photons disperse  and a leading 
fraction of them form a 
radiation front moving with the speed of light. 
The distribution of the bulk Lorentz factor smooths and adjusts to the 
radiation front (Fig. 4, middle panel) moving ahead with a near-luminal velocity. The absence of a 
viscous jump at the post-catastrophe stage means that new protons join the shock 
being cold: they are smoothly decelerated (or accelerated in the rest frame) by
radiative reaction via pairs and magnetic field.  

\begin{figure}
\centerline{\epsfig{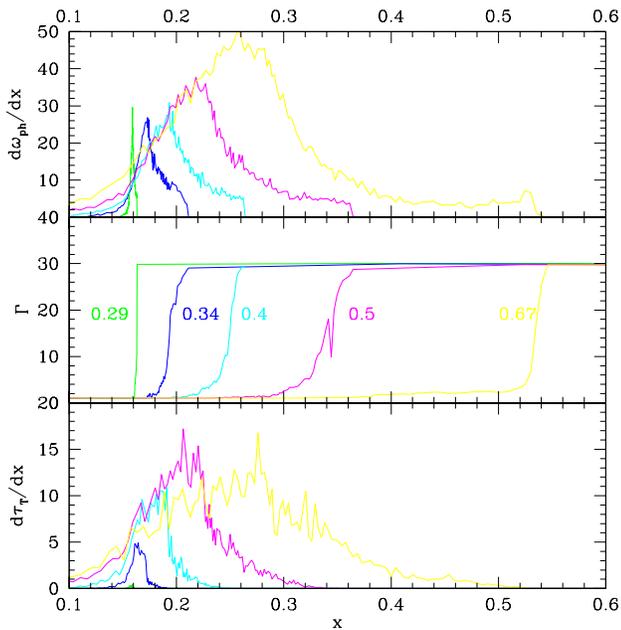}}
\caption{The distributions of the upstream photon energy density, 
the bulk Lorentz
factor and the pair number density across the shock (upper, medium and lower panels 
respectively). $x=0$ corresponds to contact discontinuity. Labels indicate the
dimensionless time since the start of the simulation. 
}
\end{figure}

 The fluid behind the main radiation front 
(where it has a semirelativistic velocity)
is not pressure balanced in this approximation. The thermal pressure of protons 
at $x < 0.15 $ swept up before the catastrophe exceeds the radiation pressure
of downstream photons. This disbalance should affect the velocity distribution 
and a real pattern can be more complicated. One can expect a semirelativistic 
forward bulk motion at the left from the broad peak of the photon distribution,
which will compress and  push pair-photon gas ahead. 

 As far as the model is very simplified and works progressively worse in attempts to 
trace the evolution further, we terminate the simulation at $t = 0.67$. At the present 
level without a careful hydrodynamic treatment we are not able to study shock 
deceleration stage and cannot reproduce the final time profile of the radiation.
We also cannot say when and how the highly radiative post-catastrophe 
state terminates.     
 
In order to represent results in terms of observed gamma-ray energy release
we have to consider different photon components in the comoving system:
 
(a). The downstream component which takes the main fraction of the 
momentum and the energy of the decelerating external medium, after the viscous
shock front disappears. Its energy is related
to the mass of the external medium $m$ swept up after the catastrophe 
as $E_d \sim m c^2 \Gamma$. 

(b). The upstream component 
(which is actually isotropic in the 
comoving frame) results from the absorption and quasi-isotropic 
reemission of the
downstream flux: $E_u \sim \eta E_d$ where $\eta$ is the average opacity 
of the system for the downstream photon flux. After the transformation 
of the upstream component to the rest (observer) frame we have:
\begin{equation}
E_{u,o} \sim \eta m c^2 \Gamma^2
\end{equation}
 
 The opacity rises during the evolution and approaches unity in the case 
of the high compactness. Indeed, all soft photons scatter in optically  
thick layer of pairs, while hard photons are absorbed due to pair production.
The deceleration of the shock can be roughly described as
\begin{equation}\label{eq:decel}
{d\Gamma \over dm} = - \eta {\Gamma^2 \over M} 
\end{equation}
  
 Where $M$ is the invariant mass (comoving energy) of the shock. Note that 
at $\eta < 1$
the deceleration is slower than in adiabatic case which is described 
in the same way as (\ref{eq:decel}) at $\eta = 1$.   

It is remarkable that the upstream spectrum 
evolves towards a GRB-like spectral shape, however still does not reach it.
The peak energy of the latest spectrum in Fig. 2 is around 70 keV. When 
blueshifting the spectrum 
to the observer frame and assuming a cosmological redshift $z = 1$ the peak energy 
comes out around 1 MeV. There exist GRBs with such peak energy, however the typical peak 
energy is less - between 200 and 300 keV. The low energy slope in this example is 
$\alpha \sim -1.5$ while for a typical GRB $\alpha \sim -1$. The values of 
the peak energy
and $\alpha$ resulting from this run are quite robust: the peak energy comes out from 
transition from Thomson to Klein-Nishina regime in Compton scattering on 
nonrelativistic or semirelativistic electrons, $\alpha = -1.5$ corresponds to 
cooling spectrum of relativistic electrons (see e.g. Ghisellini, Celotti, \& Lazatti 2000). 

Summarizing, one 
should conclude that at the present level the model does not reproduce GRB spectra.
 Nevertheless, the phenomenon provides a perspective to address the problem
due to efficient pair production and generation of optically thick layer of pairs
behind the radiation front (see Fig. 3 and Fig. 4, lower panels). Once we have an optically 
thick pair plasma, we can expect that it will produce a thermal Comptonization 
peak which may describe GRB spectra (Thompson 1997, Ghisellini \& Celotti 1999, Liang 1999). In this 
simulation such peak cannot be 
reproduced since cooled pairs are not Maxwellized:  their energy distribution is 
in Compton equilibrium with the photon spectrum. This is a consequence 
of the simplified 
approach omitting various processes of potential importance. Particularly,
the spectra could be modified in a proper direction by thermal bremsstrahlung
and double Compton scattering (for a review see Thompson 1997). 

Actually, cool 
pairs should be subject 
to a variety of collective phenomena, i.e. dissipation of plasma waves or turbulence.
As the result a prominent Comptonization peak can appear. Note, that the asymptotic case 
of the Comptonization peak is Wien distribution which is much harder in low 
energy part ($\alpha = +2$) than GRB spectra. A partially developed Wien peak 
could be a promising solution of the problem. 
   
 The contradiction in the location of the spectral break $\varepsilon_p$ 
can be relaxed if the initial Lorentz factor of the shock is 20 - 40 
rather than 100 -- 300 and 
the peak luminosity being emitted by a partially decelerated shock.

\section{Exponential Boosting of a Seed High Energy Component} 

The first example in Section 4 presents a full nonlinear simulation of the 
system without any initial seed photons. Here we show that the existence of 
a very small amount of high energy photons at the start extends the possible
range of the catastrophic evolution down to a much lower external density than 
it was probed in Section 4. The most suitable source of such photons
is Fermi acceleration of protons with photo-meson production and possible $p-n$ 
conversion suggested by Derishev et al. (2003). 
 
 Because of a high computing cost of the full nonlinear simulation 
in this section we explore the parameter space simulating the cascade  
boosting in a linear regime using linearized 
version of LPMC code. This means that the target LP field is not affected
by the simulated high energy component. 
A high boosting rate ensures that 
the system evolution eventually reaches the catastrophe stage even at a small seed 
amount of high energy photons.

 During the boosting, the spectrum of the high energy component evolves until 
it reaches 
some steady state shape 
that we do not know {\it a priori}. Therefore we have to perform a full 
simulation of the high energy component tracking its growth to a sufficiently long
time. Then, in order to handle an exponential branching of the cascade tree we have 
to use statistical weights: to cut some branches compensating this by an increase 
of the statistical weight of remaining branches. 
 The main technical problem in this simulation is large fluctuations in cascade
histories which have a very slow convergence when we accumulate the average history. Since the seed target photon field has an upper cutoff at $\varepsilon_2$
the high energy component was simulated down to the lower cutoff 
$1/\varepsilon_2$: photons below this cutoff do not produce pairs.

 Magnetic field and the seed (target) radiation field were taken according 
to \S\S 2.2 and 2.3.  The parameters of the seed spectrum were frozen
at $\alpha = -2$ and $\varepsilon_2 = 10^{-4}$. With this assumptions we reduce 
the parameter space to two dimensions:
$\Gamma$ and $\rho$. Because the assumptions are not obvious, it would be reasonable
to accept the following strategy: to explore $\Gamma, \rho$ space with the 
standard set of assumptions 
and, at one point,  to defreeze other parameters and to estimate the 
effect of their variation.

  The average breeding time profiles for $\Gamma=50$,\ $\rho = 10^4$ cm$^{-3}$
corresponding to the first three lines in Table 1 are shown in Fig. 5. 
One can see a confident exponential rise in the number of hard photons crossing the 
shock upstream.
The results on the breading rate for $\Gamma = 50$, $\rho = 10^4$ with 
different sets of other parameters 
are summarized in Table 1. 

\smallskip
\begin{tabular}{ccccc}
\hline
 &``Nonstandard'' parameters&  $S_{14}$ & $R/S$ \\
\hline
1&Standard set&5.5&21.0\\
2&$H_e = 0.2 H_e^{\rm o}$&31.&4.8\\
3&$H_e =5 H_e^{\rm o}$& 2.4&48.2\\
4&$\omega_B = 10\omega_B^{\rm o}$&37.&3.1\\
5&$\omega_B = 0.1 \omega_B^{\rm o}$&2.0 &58.8\\
6&$\alpha = -1.75$ & 2.9&40.3\\
7&$\alpha = -1.5$ & 5.1& 23.0 \\
8&$\alpha = -2.25$& 21.0& 2.5 \\
9&$10^{-8} < \varepsilon < 10^{-3}$&4.7& 25.0\\
10&$10^{-8} <\varepsilon < 10^{-5}$&15.3& 7.6\\
\hline
\end{tabular}
\vskip0.1cm
{\small{\bf Table1} The boosting rate of the high energy component
for $\Gamma = 50$, $\rho=10^4$ ($\ell_{\rm o} = 7.3$) depending on other parameters.
Second column gives those parameters which differ from the standard 
set (see text). Third column shows the boosting rate in terms of the
distance of the shock propagation $S$ where the high energy component grows by 
an order of magnitude$S$.
 Fourth column gives the total growth of the 
high energy component (orders of magnitude) when the shock propagates 
the distance 
$R = 0.5 R_d = 1.17 \cdot 10^{16}$cm, where the deceleration 
distance $R_d$ implies the 
initial isotropic kinetic energy $10^{54}$ erg. The standard parameters are:
$\ell_e = 3.5 \cdot 10^{-3}$,  $H^{\rm o} = 14.7$ G, 
$\omega^{\rm o}_B = 1.4 \cdot 10^{-3}$, $H^{\rm o}_e = 10^{-3}$G, $\varepsilon_2 = 10^{04}$, $\alpha = -2$  
}  
 
\smallskip

 First of all it should be noticed that the breeding rate at moderate parameters 
$\Gamma = 50, \rho=10^4$cm$^{-3}$ is in a large excess over required to 
trigger the catastrophe in the most of trials.
One can see a strong dependence on magnetic field. The effects of $H$ and $H_e$ variation are of opposite 
signs as it was qualitatively shown in \S 2.4. 

\begin{figure}
\centerline{\epsfig{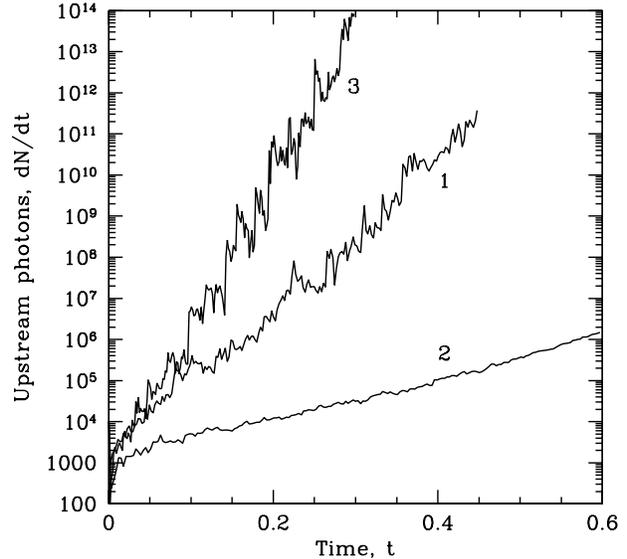}}
\caption{ Average breeding time histories for $\Gamma = 50$, $\rho = 10^4$cm
and different external magnetic field. Labels indicate corresponding line 
numbers in Table 1.
}
\end{figure}

\begin{figure}
\centerline{\epsfig{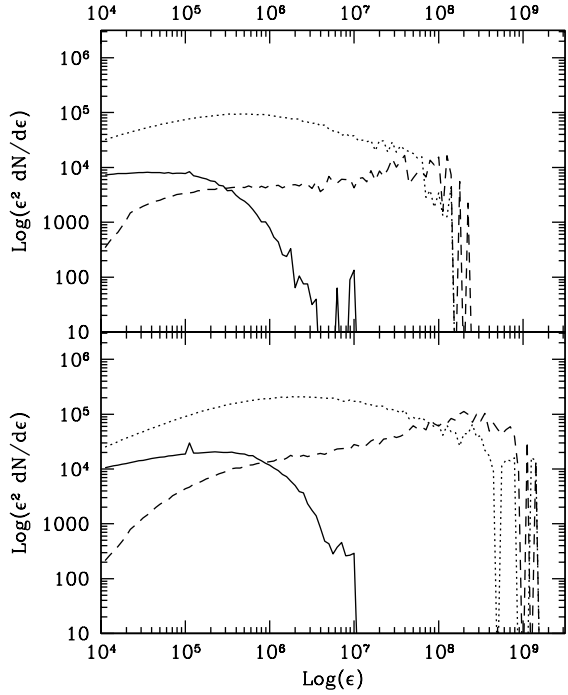}}
\caption{
Spectra of particles crossing the shock for 
$\Gamma = 50$, $\rho = 10^4$cm$^{-3}$
(upper panel) and $\Gamma = 50$, $\rho = 10^3$cm$^{-3}$ (lower panel) 
and standard set of other parameters. 
Solid lines: upstream photons; dotted lines: downstream photons; dashed
lines: downstream pairs. Spikes at $\varepsilon = 10^5$ represents 
monochromatic photons injected at $t=0$.
}
\end{figure}

Figure 6 demonstrates spectra of particles crossing the shock (energy of all particles is 
in the comoving system).
One can see that the main fraction of energy is transferred downstream across the shock by 
high energy Comptonized photons rather then by advected pairs.  
The turnover in spectra of upstream photons qualitatively agrees with equation (\ref{eq:gmax}) yielding
$\gamma_{\rm max} \sim 3 \cdot 10^6$  for $\rho = 10^4$ and $\gamma_{\rm max} \sim 0.7\ 10^7$ for 
$\rho = 10^3$. A larger difference in these two cases visible in Fig. 6 appears due to 
a higher pair production opacity for high energy photons in the first case. 

 Fig. 7 demonstrates the map of $\Gamma - \rho$ space when other parameters are frozen as 
described above. Levels of the constant compactness and the constant time scale are given for 
total isotropic kinetic energy $E = 10^{54}$ erg.  Both compactness and the time scale 
vary as $E^{1/3}$ in the spherical geometry and at the constant density.  The 
broken solid line 
shows the catastrophe threshold which is conventionally defined as the condition that the
cascade is boosted at least by 10 orders of magnitude when the shock propagates the distance 
$R = 0.5 R_d$. Perhaps one could use a weaker requirement: we prefer to use this conservative 
criterion since the real intensity of the seed high energy radiation is unknown.   
 
\begin{figure}
\centerline{\epsfig{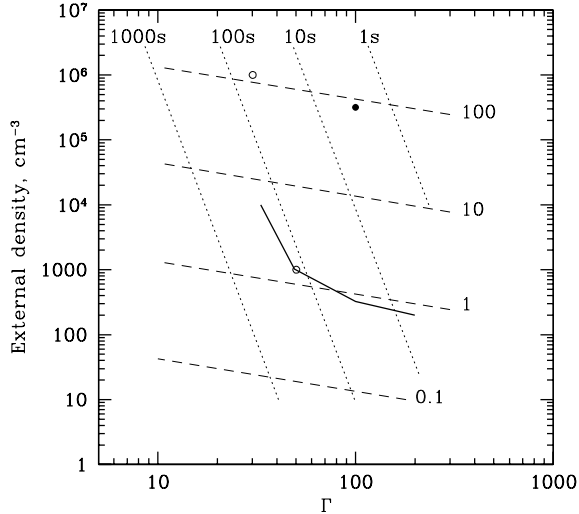}}
\caption{ The explored area in $\Gamma - \rho$ coordinates
(other parameters are frozen as described in the text).
The solid line shows the threshold for the catastrophe
estimated with the conventional condition that the seed high energy component 
grows at least by 10 orders of 
magnitude as the shock propagates the distance $R = 0.5 R_d$. 
Dashed lines represent the levels of the constant observer time scale $R/(\Gamma^2 c)$ and dotted lines correspond to 
the constant comoving compactness $\ell_{\rm o}$ assuming $E = 10^{54}$ erg. Circles represent 
parameters for full nonlinear simulation runs described in Sections 4 and 6.
}
\end{figure}

 The threshold line in Fig. 7 depends on the frozen parameters, i.e. it is model dependent.
As it was mentioned in \S 2.4 the strong slowly dissipating magnetic field in the shock would
substantially rise the threshold. On the other hand if the field dissipates very fast leaving 
a dominating large scale component, the threshold would be lower.  
The range of the catastrophe would be extended if an external field stronger than 
assumed is induced due to spherically asymmetric Compton drag as 
discussed in \S 2.2.

\section{GeV bursts}

 Two runs described in Section 4 correspond to a comparatively high compactness parameter
when the main energy release occurs in the soft gamma range. In Section 5 it is shown that the 
catastrophe is possible at a much less compactness (see Fig. 7). In such case we can expect 
to observe a much harder radiation which would approximately correspond to early spectra 
in Fig.1. In order to demonstrate this directly we performed a run for $\Gamma = 50$, 
$\rho = 10^3$cm$^{-3}$. The compactness is $\ell_{\rm o} = 1.1$, 
``standard'' magnetic field in the 
shock $H=4.4$ G, characteristic radius $R = 2.0 \cdot 10^{16}$cm,
observer time scale $T_{\rm o} = 133$ s. The ``minimal model'' does not work at such 
parameters as the primary synchrotron radiation is strongly self-absorbed, therefore the same 
standard ad hoc hypothesis of the seed radiation as in Section 5 has been 
used: $\ell_s = \ell_{\rm o} m_e/m_p$,
$10^{-8} < \varepsilon < 10^{-4}$, $\alpha = -2$. The catastrophe at such parameters 
develops due to the breeding of external high energy photons. It was assumed that 
the high energy component due to preceding breeding has gained the luminosity 
$\omega_h = 0.005 \omega_s \sim 2.5 \cdot 10^{-6} \ell_{\rm o}$. The corresponding amount of upstream
high energy photons with $\varepsilon = 10^5$, has been injected at the start of the simulation.

\begin{figure}
\centerline{\epsfig{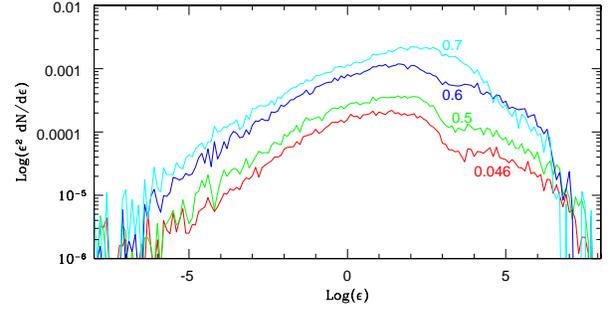}}
\caption{Instantaneous spectra of photons for $\Gamma = 50$, $\rho = 10^3$cm$^{-3}$.
Labels 
indicate the dimensionless time $t$ from the start of the simulation. Time unit corresponds 
to 133 s in the observer frame.}
\end{figure}

 The resulting sequence of instantaneous upstream photon spectra is shown in Fig. 8.
 Generally, at a low compactness we can expect a maximum
of $\varepsilon^2 dN/d\varepsilon $ distribution in high energies and an 
approximate power law spectrum
in a wide range with the index between $\alpha = -1.5$ (fast cooling spectrum) 
and $\alpha = -2$ (saturated pair cascade, Svensson, 1987 ). In this case the maximum energy 
release is at the rest frame energy $\sim 5$ GeV (the spectrum in Fig 8 is not final,
however, since some hardest photons will be absorbed on the way to the observer).

 According to Fig. 7, the catastrophe threshold at $\Gamma \sim 50$ is $\ell_{\rm o} \sim 1$ when the
main energy is still released in hard gamma-rays. This value is
model dependent, the threshold can be higher for less favorable combination of parameters,
e.g. for higher $\omega_B / \omega_s$ ratio. Nevertheless, if the environment of GRB 
progenitors is diverse, we can expect that GeV bursts should be as usual phenomenon as 
classic GRBs.  On the other hand, they are more difficult for observations. 
Such events should be weaker in soft gamma range (soft gamma fluence is one  -- two orders 
of magnitude less than the high energy fluence). And, more importantly, they should be longer 
and smoother (see Figs. 7 and 9). The example of this section corresponds to 
characteristic
time scale 133 s (the total duration can be longer). Such events are more difficult for a triggering  
and an off-line search than classic GRBs, unless 
an event is strong enough.
The long duration also makes difficult a time-location cluster analysis for high energy photons
in existing data.  

\begin{figure}
\centerline{\epsfig{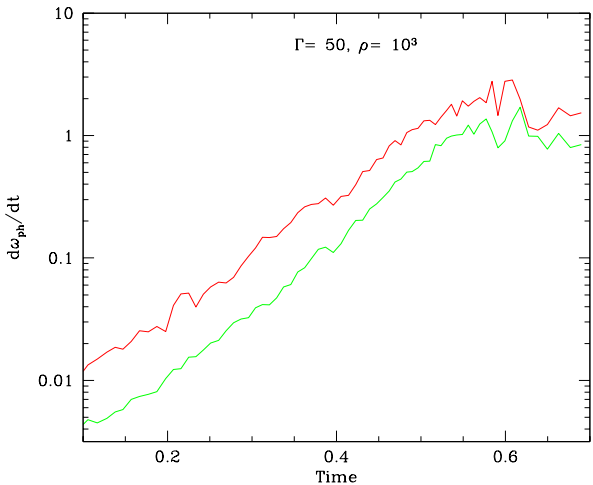}}
\caption{The evolution of the electromagnetic component with time for 
$\Gamma = 50$, $\rho = 10^3$cm$^{-3}$.
The time derivative of the total energy of upstream (lower curve)
and downstream (upper curve) photons is shown.
}
\end{figure}

 Has something like this been observed? A candidate for such a GeV burst is GRB 970417a 
detected simultaneously by BATSE and Milagrito (Atkins et al. 2000) in the range above 
50 GeV.
The inferred fluence above 50 GeV is at least an order of magnitude greater than
the fluence in the BATSE range which corresponds to the type of the spectrum shown in Fig. 8 if the maximum is at an order of magnitude higher energy. 
However, it was a comparatively short burst with $T_{90} = 7.9$ s. Then following Fig. 7 one 
has to assume a high Lorenz factor ($\Gamma \sim 200$), or a less kinetic energy at 
a lower $\Gamma$. Note that there should be a strong selection effect in favor of short
GeV bursts.

 Another hint at GeV bursts provide EGRET data. The famous GRB 940217 (Hurley et al. 1994) where EGRET
has detected a delayed GeV emission during an hour after BATSE trigger hardly can 
match the concept of a GeV burst. It rather looks like a superposition of a classic 
GRB and a GeV burst. In principle this is possible when the shock meets  
localized 
dense clouds while propagating in a less dense environment. Better candidates for GeV bursts can
be several EGRET detected GRBs reported by Catelly, Dingus \& Schneid (1997) and Dingus, 
Catelly \& Schneid
(1997). Some of them have high energy spectral indices $\alpha \sim -2$ up to GeV range.
 Eventually the issue of GeV bursts will be clarified by GLAST mission.

\section{Discussion} 

 While the effect of the electromagnetic catastrophe can, in principle, be applied 
to different astrophysical phenomena, its application to GRBs is of the highest interest.
Below we discuss the main issues where the effect can be relevant. 

\subsection{The radiative efficiency and the GRB scenario}

The considered effect straightforwardly provides the ultimate ($\sim$ 100\% 
radiation efficiency
and, in this respect, it probably has no competing mechanisms.
The phenomenon of GRBs does require a high radiation efficiency. The total isotropic
soft gamma ray energetics of GRBs with measured redshift varies between 
$10^{52}$ and $5 \cdot 10^{54}$ erg. A low efficiency would imply much 
higher values for the isotropic kinetic energy of the ejecta which 
probably contradicts the afterglow data.

 The catastrophe is a model independent effect which, however, has a set of threshold 
conditions. If GRBs were associated with coalescence of neuron star binaries,
the effect would never work since the expected density
of the environment is too low: 0.1 -- 1 cm$^{-3}$. However a wealth of 
recent data support the collapsar scenario where we can expect a much higher 
density and the phenomenon may occur (see M\'esz\'aros 2002 for a review). 
Is it possible that the threshold conditions
are still never satisfied despite a sufficient density of the environment?
Formally, it could be if the external magnetic field is always purely radial.
But as is was discussed in \S 2.2 the radial field structure can hardly survive 
the early stage of a GRB in any scenario. Other inhibiting factor could be 
a strong domination of magnetic energy density over radiation energy density.
However, we do observe GRBs implying a high radiation density in the source.
Whatever produces this radiation at the start should end up with the 
electromagnetic catastrophe.  

 The effect is associated mainly with external shock scenario as internal 
shocks have too low relative Lorentz factor. On the other hand, internal shocks 
can radiate only a relatively small fraction of the total kinetic energy because of
kinematic constraints. A competition with a much more efficient mechanism presents
a new difficulty for the internal shock scenario. 
For example, let us consider the typical internal shock scenario:
prompt GRB emission from $R \sim 10^{13} - 10^{14}$ cm due to collisions of multiple shocks 
and the following afterglow due 
to an external shock at larger radii. An important modification to this scenario 
introduces the effect discovered by Beloborodov (2002): 
the radiation emitted by internal 
shocks sweeps out the external matter up to radius 
$R \sim 10^{15} - 10^{16}$cm leaving an empty cavity. Now let us consider
the situation when the shock reaches the edge of the cavity. 
It meets the environment with enhanced density and optical depth (since
the swept out matter is compressed and loaded by a large amount of pairs), 
with a non-radial magnetic field (the radial field structure can hardly 
survive such event), with intense seed radiation (side scattered photons).
These are perfect conditions to give rise to the electromagnetic catastrophe
with a proper GRB time scale.
If so, the radiative energy release in such collision would outshine
a burst from internal shocks by 1.5 -- 2 orders of magnitude 
(the main fraction of total kinetic energy of the ejecta at the full radiative 
efficiency versus a few per cent of the kinetic energy at unknown radiative
efficiency).
This would look as the main event, rather than the afterglow. Even if the main 
energy release is in the GeV range (low compactness, see Section 6), such events can hardly 
escape systematic observations.  

\subsection{Issue of the time variability}

The internal shock model is motivated mainly by difficulties with description of the 
GRB time variability arising in the external shock scenario (Fenimore 1996).
Dermer (2000), on the other hand, suggests a number of possible solutions for external shocks.
The effect of the electromagnetic catastrophe may extend the list of possible
solutions.  

 1. The catastrophe is a nonlinear effect which can amplify any fluctuations in 
external density and seed radiation. In addition it is sensitive to the geometry 
of magnetic field.
One dimensional treatment performed in this work 
loses many important features which 
could appear in a 3D simulation with realistic hydrodynamic treatment. The catastrophe most 
probably will develop locally in many spots within $1/\Gamma$ cone at different time.
The spherical symmetry of the shock should be distorted due to a strong feedback between 
the radiation and the fluid pattern. As a result one can observe a complex time 
structure even at an approximate symmetry in the initial state.

 2. The post-catastrophe evolution was interpreted here as a quasi-steady state. Actually, 
especially in the case
of decreasing density, it could be unstable and recurrent. This could be, e.g., due to a version  
of the Beloborodov effect: the radiation of external shock passing through the dense matter 
sweeps out a less dense matter ahead, radiation decays until the shock catches up 
the swept matter. Then the shock regenerates producing a new emission episode. 
In this way one can explain episodes separated by long quiescent intervals.

 3. The spherical symmetry of ejecta can be completely broken prior the emission stage
up to fragmentation into a bunch of separate droplets of different transversal 
size $X_T < R/\Gamma$ (this implies a hydrodynamic transversal confinement of droplets
which needs a separate study).
This require a strong instability (most probably of Rayleigh-Taylor type) 
at some stage. A model representing a GRB ejecta as a shower of 
blobs was proposed by Heinz \& Begelman (1999). This is the most attractive 
possibility because in addition to the complex time structure it can produce a chain 
reaction: the catastrophe in one droplet supply seed photons to a neighboring 
droplet and induce the next catastrophe. It was shown by Stern \& Svensson (1996)
that the chain reaction gives a  statistical and 
qualitative description of GRBs temporal diversity. 

 All these suggestions require 3D numerical studies. 

\subsection{Problem of the spectral break}  

 The observed spectral break in the sub-MeV range with a very hard low energy 
slope is still a serious problem. It seems that this problem
is common for all GRB models assuming a high Lorentz factor and a small optical depth.
Two alternative explanations of the spectral break for optically thin 
emitting medium were suggested: optically thin synchrotron emission in slow cooling 
regime
and synchrotron self-absorption break.  For the criticism of the former see Ghisellini, Celotti \& 
Lazatti (2000). The latter implies too high magnetic field (in our case the self-absorption break 
should appear at $\varepsilon \sim 10^{-6} - 10^{-5}$ in observer frame). 

As it was discussed in Section 4 the electromagnetic catastrophe at a high compactness provides 
an optically thick pair loading and thus gives a perspective to explain the break with thermal
Comptonization peak. However there remains the issue whether the thermal Comptonization is 
sufficiently fast and efficient. If it works, then we should find an anticorrelation between 
the sharpness of the break (implies high compactness) and the high energy emission (implies low 
compactness). If the break is still too persistent then we have to conclude that
our consideration misses some important detail. For example, a possible role of ejecta behind the
contact discontinuity was ignored while actually it could be an efficient photon reflector.
 
 The location of the break hints at a comparatively low Lorentz factor and therefore a high density
and small deceleration radius. It could be that the range of parameters studied in this work is
actually far from the typical GRB ``working regime''.

\subsection{Summary}

 The effect of the electromagnetic catastrophe outlined here in the ``zero approximation'' 
can explain the high radiative efficiency of ultrarelativistic shocks while the explanation
of the GRB time variability and spectra still require a lot of work. The study of the time 
variability should certainly rely on a detailed three dimensional hydrodynamics. The 
understanding of the spectra requires a more complete description of the shock structure
and physics of particle interactions for nonrelativistic and semirelativistic pairs.

\section*{ACKNOWLEDGMENTS}

The author thanks K. Hurley, A. Illarionov, P. Ivanov,
J. Poutanen for useful discussions and Y. Tikhomirova for assistance.
This work was supported by the Russian Foundation for 
Basic Research (grant 00-02-16135). A part of this work was done during a 
visit to the University of  Oulu with the support of the grant 
from the Jenny and Antti Wihuri foundation.




\begin{thebibliography}{99}

\bibitem[ ]{a1}
Atkins R., et al, 2000, ApJ, 533, L119

\bibitem[Ostr]{b1}
Bednarz J., Ostrowski M., 1999, MNRAS, 310, L11

\bibitem[Beloborodov]{b2}
Beloborodov A.F., 2002, ApJ 565, 808.

\bibitem[Blandford]{b3}
Blandford R.D., McKee C.F., 1976 Phys Fluids 19, 1130

\bibitem[ ]{c1}
Catelli J. R., Dingus B. L., Schneid, E. J, 1998, 
in Meegan C.A., Preece R.D.,
Koshut T.M., eds, AIP conference proceedings, 428,.
Gamma-Ray Bursts : 4th Huntsville Symposium, Huntsville,  p 309


\bibitem[CR]{c2}
Cavallo G, Rees M.J., 1978, MNRAS, 183, 359

\bibitem[Derishev]{d1}
Derishev E.V., Aharonian F.A., Kocharovsky V.V., Kocharovsky Vl.V., 2003,
astro-ph/0301263

\bibitem[Dermer]{d2}
Dermer C.D., 2000, Astro-ph/0001389

\bibitem[ ]{d3}
Dingus B. L., Catelli J. R., Schneid, E. J., 1998,
in Meegan C.A., Preece R.D.,
Koshut T.M., eds, AIP conference proceedings, 428,.
Gamma-Ray Bursts : 4th Huntsville Symposium, Huntsville,  p 349

\bibitem[Fenimore]{f1}
Fenimore E., Madras C.D., Nayakshin S., 1996, ApJ, 473, 998

\bibitem[Ghisellini1]{g1}
Ghisellini G., Celotti A., Lazatti D., 2000, MNRAS, 313, L1

\bibitem[Ghisellini2]{g2}
Ghisellini G., Celotti A. 1999, A\&AS 138, 527

\bibitem[ ]{hb}
Heinz S., Begelman M.C., 2000, ApJ, 535, 104 

\bibitem[ ]{h1}
Hurley K., et al. 1994, Nature, 372, 652 

\bibitem[Kirk]{k1}
Kirk J.G., 1997, 
in Ostrovski M., Sikora M., 
Madejski G., Begelman M., eds,
 Relativistic Jets in AGNs, Proceedings of the 
International Conference, Kracow p. 145 

\bibitem[Liang]{l1}
Liang E., 1999, A\&AS, 138, 529

\bibitem[Medvedev \& Loeb]{mb}
Medvedev M.V., Loeb A., 1999, ApJ, 526, 697   

\bibitem[me]{mz}
M\'esz\'aros P., 2002, ARA\&A, 40, 137

\bibitem[Meszaros]{m1}
M\'esz\'aros P., Ramires-Ruiz E., Rees M.J., 2001, ApJ 554, 660

\bibitem[RR]{r1}
Ramires-Ruis E., Madau  P., Rees M.J., Thompson C., 2001,
in Costa E., Frontera E., Hjorth J., eds, 
Gamma-Ray Bursts in the Afterglow Era, Proceedings of the International workshop, Rome, Berlin Heidelberg: Springer, p. 278.
.
\bibitem[Rees]{r2}
Rees M.J., M\'esz\'aros P., 1992, MNRAS, 258:L41

\bibitem[]{s1}
Stern B.E., 1985 Astron Zh+, 62, 529

\bibitem[Stern]{s2}
Stern B.E., Begelman M.C., Sikora M., Svensson R., 1995, MNRAS, 272, 291 

\bibitem[]{s3}
Stern B.E., Svensson R., 1996, ApJ, 469, L109

\bibitem[Svensson]{s4}
Svensson R., 1987, MNRAS, 227, 403

\bibitem[Thompson]{t1}
Thompson C., Madau P.  2000, ApJ 538, 105

\bibitem[Thompson1]{t2}
Thompson C., 1997, 
in Ostrovski M., Sikora M., 
Madejski G., Begelman M., eds,
 Relativistic Jets in AGNs, Proceedings of the 
International Conference, Kracow, p. 63


\end{thebibliography}
\end{document}